# SELECTED ASPECTS OF THE CONDENSED MATTER PHYSICS. WORK AND LIFE OF PROFESSOR B.YA. SUKHAREVSKY

V.V. Eremenko [1], V.G. Manzhely [1], V.N. Varyukhin [2], A.D. Alekseev [3], V.A. Beloshenko [2], A. Voronel [3], V.V Pustovalov [1], V.Ya. Maleev [5], S.A. Gredeskul [6], L.P. Mezhov-Deglin [7], A.Yu. Zakharov [8], N.Ya. Fogel [9], I.I. Vishnevsky [10], V.M. Tsukernik [11], V.N. Vasyukov [12], E.P. Feldman [3], A.V. Leont'eva [13], Yu.A. Mamaluy [14], G.G. Levchenko [2], Yu.V. Medvedev [2], A.D. Prokhorov [2], V.M. Yurchenko [2], B.G. Alapin [15], V.G. Ksenofontov [16], A.M. Bykov [17], E. Lakin [9], T.L. Pyatigorskaya [13], S.V. Lysak [18], A.I. Erenburg [6], S. Kovalenko [1], G.E. Shatalova [2], E.O.Tsybul'sky [2], A.Yu. Prokhorov [2], I.V. Zhikharev [19]

[1] B.I. Verkin Institute for Low Temperature Physics and Engineering, Kharkov, 61103, Ukraine
[2] A.A. Galkin Donetsk Institute for Physics and Engineering, Donetsk, 83114, Ukraine
[3] Institute for Mining processes physics, Donetsk, 83114, Ukraine
[4] Tel-Aviv University, P.O. Box 39040, Tel-Aviv 69978, Israel
[5] A.Ya. Usikov Institute of Radiophysics and Electronics, Kharkov, 61085, Ukraine
[6] Ben-Gurion University, Beer-Sheva, Israel
[7] Institute for Solid State Physics, Chernogolovka, Moscow region, 142432, Russia
[8] Novgorod State University, Novgorod the Great, 173003, Russia
[9] Technion, Haifa 32100, Israel
[10] New York, USA
[11] Rehovot, Israel
[12] GasProm, PromGas, Moscow, Russia
[13] Haifa, Israel
[14] Donetsk National University, Donetsk, 83055, Ukraine
[15] Koblenz, Germany
[16] Mainz, Germany
[17] New Jersey, USA
[18] Kharkov, Ukraine
[19] Lugansk National University, Lugansk, Ukraine

The paper contains a short description of the scientific papers of Professor Boris Ya. Sukharevsky on condensed matter physics. His basic scientific activity was related to following directions: structural vacancies concept; influence of point defects on statics and kinetics of phase transitions; DNA crystallization and low-dimensional systems properties; phase transitions in Jahn-Teller crystals; Thermophysics of superconductors e.a. The list of most essential papers of Prof. B.Ya. Sukharevsky is included.
16 pages, 53 references.

# К восьмидесятилетию профессора Бориса Яковлевича Сухаревского

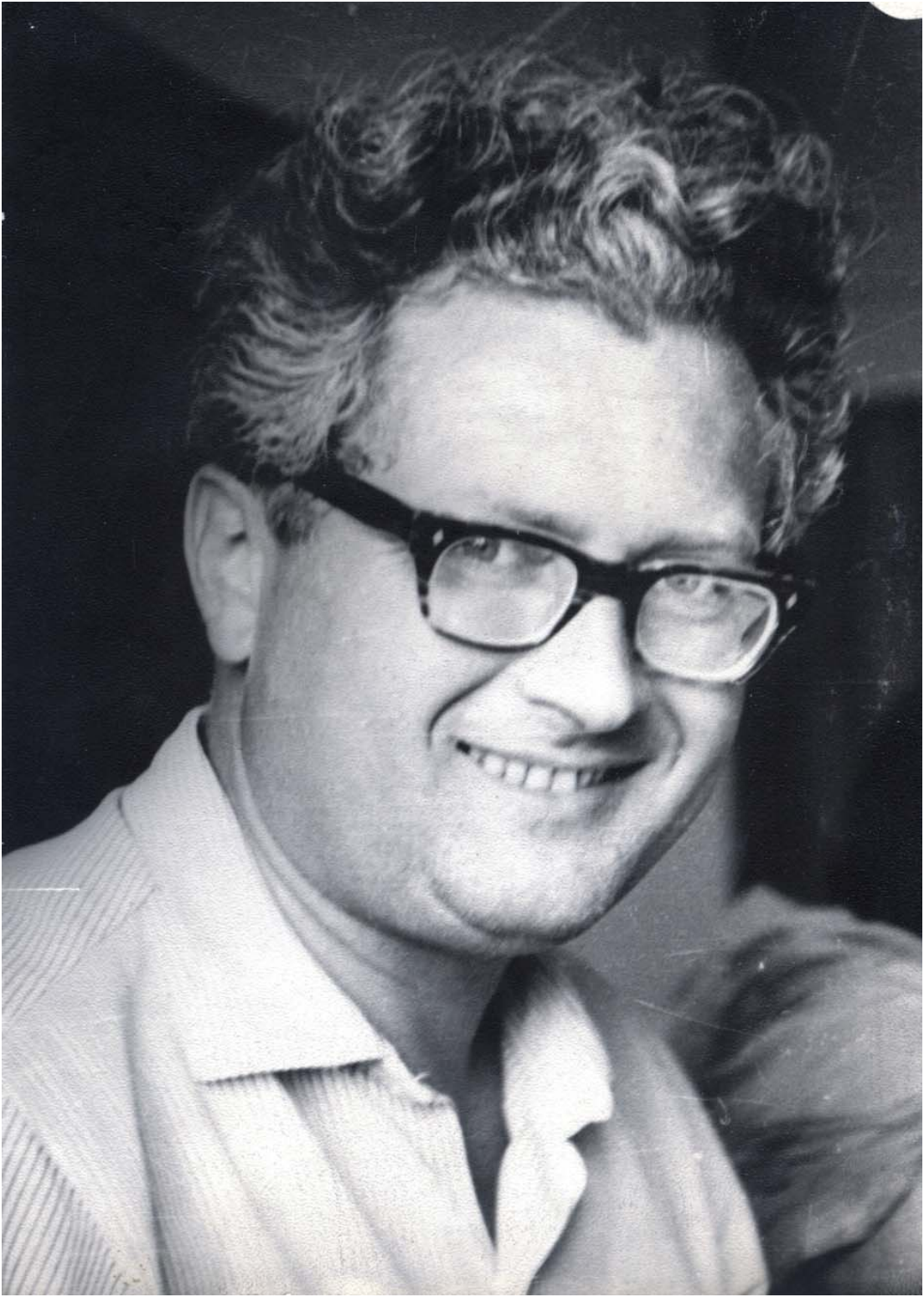

**1930–2001**

Известный советский ученый профессор Борис Яковлевич Сухаревский родился 15-го августа 1930 года в гор. Житомире в семье кадрового военного. Борис Яковлевич (Б.Я.) закончил кафедру Физики твердого тела Физико-математического факультета ХГУ и, с легкой руки руководителя кафедрой профессора Пинеса, тоже Бориса Яковлевича, избрал рентгеноструктурный анализ одним из главных методов научных исследований в своей экспериментальной деятельности. Его научная карьера началась в 1953 году в харьковском Институте огнеупоров, в физической лаборатории, созданной опять-таки профессором Пинесом в конце 30-х годов. Несмотря на то, что институт был отраслевой (не академический), лаборатория была хорошо оснащена приборами для исследований, а главное, в ней работали первоклассные физики и инженеры! Где-то в 55-м году разразился скандал из-за острофокусной рентгеновской трубки. Б.Я. Пинес (с его авторитетом!) заявил, что его трубка (Пинеса–Безверхого) имеет лучшие параметры и более острый фокус, чем трубка Дудавского–Чупринина в Огнеупорах. Поскольку Сухаревский работал на обеих трубках (студентом на кафедре Пинеса, а затем в Огнеупорах), то его и сделали председателем комиссии по проверке: чья же трубка лучше. Святая простота сильно подвела Б.Я., когда он на заседании комиссии признал лучшей... трубку Федора Ивановича Чупринина! Пинес тогда разошелся не на шутку, заявив, что жалеет, что Сухаревский был его Учеником! Размах его возмущения с его огромной, видимо, энергетикой... Словом, через несколько дней у Сухаревского случился рентгеновский ожог левой руки (Б.Я.-левша и юстировал образец прямо под пучком) на той же трубке Дудавского–Чупринина. Вот такое совпадение! Полгода врачи спасали руку. Вроде помогла мазь Вишневского, еще и потому, что Б.Я. подружился в то время (и на всю жизнь!) с Илюшей, тоже Вишневским!

Надо сказать, что огнеупорные материалы, несмотря на сложность строения, оказались целым кладезем научных открытий. По мнению учеников Б.Я., в 50–60-е годы серьезных работ по физике огнеупоров и керамики почти не было. И Б.Я., имея за плечами школу Б.Я. Пинеса и Я.Е. Гегузина, а также несомненный Божий дар, сумел за какие-то 12 лет работы в институте стать одним из ведущих специалистов в стране. Авторитет его на конференциях всех уровней был огромен. И неверно было бы сказать, что он продолжил чьи-то традиции. Б.Я. был фактически одним из основателей новой области физики — физики огнеупоров. Вместе с И.И. Вишневским он построил теорию окислительно-восстановительных процессов в оксидах с кубической решеткой, объяснил парадоксальную температурную зависимость теплопроводности сложных шпинелидов, вызывающую саморазогрев в стенке металлургического агрегата. Впрочем, всего не перечислить.

Когда через десять лет работы у Б.Я. накопился огромный материал интереснейших результатов, сотрудники, друзья (и жена дома!) в один голос заговорили о защите его кандидатской диссертации. Но где взять руководителя? (Кстати, кроме Б.Я. Пинеса — руководителя диплома — у Б.Я. никогда более научных руководителей не было. Но обратиться к Пинесу за помощью тогда Б.Я откровенно боялся, т.к. хорошо помнил и скандал, и ожог!) Руководителем, пожалев Б.Я., стал профессор Лев Самойлович Палатник, отметив, кстати, что «материала здесь на две диссертации». Защиту во ФТИНТе помогал организовать большой друг Б.Я. Виталий Валентинович Пустовалов (к тому времени он уже перешел во ФТИНТ из Огнеупоров). Они оба защищались на одном совете. Среди отзывов на диссертацию Б.Я. красовался блистательный отзыв профессора Б.Я. Пинеса!

С 1965 года Б.Я. Сухаревский работал уже во ФТИНТе Академии наук УССР, с 1970 года — руководителем Отдела биофизики, а с 1978 года — руководителем Отдела фазовых переходов в Донецком физико-техническом Институте Академии наук Украины. С 1992 по 1996 г.г. был зам. директора института по науке. Одновременно он преподавал в Харьковском, а затем в Донецком Госуниверситетах.

Круг проблем, которыми занимался Б.Я Сухаревский, достаточно широк: физика твердого тела, физика низких температур, кристаллография, биофизика, сверхпроводимость, физика фазовых переходов. Во всех этих областях Б.Я. Сухаревскому принадлежат поистине пионерские работы. Великолепный экспериментатор, он обладал редкой научной интуицией в постановке исследований и, к тому же, таланливый теоретик, с легкостью обсчитывал свои результаты и создавал грамотные научные концепции. Когда-то академик Веркин, возглавляя какую-то плановую комиссию в ДонФТИ, назвал Б.Я. своим «талантливым учеником», который свой у экспериментаторов и свой у теоретиков!

В его многолетней научной деятельности, отраженной в 283 научных работах, можно отметить следующие успешно решенные проблемы:

## 1. Разработка концепции структурных вакансий

Б.Я. Сухаревский одним из первых ввел в рассмотрение структурные вакансии в координационных кристаллах, возникающие при иновалентном замещении и при изменении валентности ионов переходных металлов. В связи с этим концентрация структурных вакансий может существенно превышать концентрацию тепловых, так что структурные вакансии представляют собой своеобразный компонент твердых растворов. Используя представление о структурных вакансиях, Б.Я. Сухаревский предложил и обосновал механизм образования твердых растворов замещения–вычитания между неизоморфными координационными кристаллами, разработал методику расчета кривых растворимости и температурной зависимости концентрации вакансий в системах, содержащих ионы переменной валентности.

## 2. Изучение роли структурных вакансий в формировании физических и, прежде всего, кинетических свойств

Было показано, что вакансии значительно эффективнее рассеивают фононы, чем другие точечные дефекты. Это позволяет регулировать теплопроводность оксидной керамики, что имеет и практическое значение. Было впервые использовано представление о комплексах вакансия–примесный ион для объяснения особенностей электро- и массопереноса в нестехиометрических координационных кристаллах.

## 3. Изучение влияния точечных дефектов – примесных атомов и вакансий – на температуру и кинетику фазовых переходов

В этой серии экспериментальных работ приведены уже и теоретические оценки, позволяющие объяснить природу «стабилизации» кубической фазы диоксида циркония. Впервые был предложен механизм зародышеобразования, учитывающий перераспределение точечных дефектов между матрицей и зародышем в процессе его роста. Этот механизм обеспечивает в пространстве «работа образования зародыша — его размер — его состав» обходную траекторию по наименьшему перевалу.

В ряде систем со шпинельной структурой было установлено существование критической концентраци структурных вакансий, превышение которой, независимо от температуры, приводит к распаду твёрдого раствора. Столь значительное влияние вакансий на энергетику кристалла приводит к сильной зависимости температуры структурных переходов от их концентрации. Это показано на примерах электронного упорядочения в магнетите и полиморфных переходов в диоксиде циркония и кристобалите. Приведенные результаты являются фундаментальными, но из них следует и важный практический вывод: существенное размытие фазового перехода и сопряженного с ним скачка объема, понижающего термическую устойчивость диоксид-циркониевой и динасовой керамики даже при небольшой преднамеренно созданной неоднородности распределения в ней структурных вакансий. При изучении полиморфизма диоксида циркония Б.Я. с учениками одним из первых установил мартенситный характер фазового перехода, что впоследствии позволило понять природу свойств разработанной в США «керамической стали».

**4. Теплофизические свойства технически важных сверхпроводников**

Эти данные необходимы для расчета теплового режима сверхпроводящих магнитных систем на самых ранних этапах их создания. В работах Б.Я. Сухаревского из температурных зависимостей теплоемкости и теплопроводности сплавов ниобий–титан и ниобий–цирконий помимо теплофизических характеристик извлекается температурная зависимость сверхпроводящей щели. Было установлено, что щель, равно как и критические поля и токи, занижена в меру содержания в сплавах кислорода, попадающего в них благодаря геттерным свойствам титана и циркония. Это открывало путь существенного улучшения качества технически важных сверхпроводящих сплавов, который, к сожалению, в 70-е годы, когда проводились данные исследования, не был реализован. Только в 90-е годы в мире начаты были работы по глубокой очистке сверхпроводящих сплавов от кислорода.

**5. Работы по кристаллизации ДНК, динамика и кинетические свойства низкоразмерных систем**

Эти уникальные исследования были проведены по предложению и инициативе академика Б.И. Веркина в связи, как это ни удивительно, тоже с поисками высокотемпературной сверхпроводимости. (Идеи Литтла, затем с середины 70-х годов — бум, вызванный обнаруженной аномалией сверхпроводимости в квазиодномерных комплексах с переносом заряда.)

Высокотемпературная сверхпроводимость на том этапе обнаружена не была, зато Б.Я. Сухаревский с соавторами, разработав методику получения «кристаллов» из нативных фрагментов ДНК, по сути, открыл новый тип упорядочения:

«кристаллы» обладали спайностью и оптической активностью, давали четкую дифракционную картину на малых углах и широкие гало на средних углах рентгенограммы, отвечающих расстояниям между нуклеотидами. Другими словами, кристаллы ДНК истинны на уровне элементов структуры — фрагментов ДНК, но каждый фрагмент несет свою наследственную информацию, имеет свою последовательность нуклеотидов. Авторы предполагали использовать методику кристаллизации для получения кристаллов, составленных из идентичных фрагментов, полученных путем копирования заданного фрагмента ДНК, который играет определенную функциональную роль в биологической системе. Это позволило бы не только установить нуклеотидную последовательность такого фрагмента, но и получить его пространственную структуру. К сожалению, уровень развития генной инженерии тогда не позволил решить задачу копирования в объеме, достаточном для получения кристалла. (Следует отметить, судя по литературе, что после этих пионерских работ Б.Я. Сухаревского и его соавторов, данную задачу успешно решили американские исследователи в 1981г.)

Начиная с работ Ильи Михайловича Лифшица, предпринимались многочисленные попытки обнаружить особенности колебательного спектра низкоразмерных систем по температурной зависимости теплоемкости. В работах Б.Я. Сухаревского использованы методы «трехмеризаций» спектра путем интеркалирования и сжатия в направлении нормали к плоскостям слоистых кристаллов. Первый способ привел к проявлению «трехмеризаций» в низкотемпературной теплоемкости, второй — в теплопроводности.

Эти работы Б.Я. Сухаревского по кристаллизации ДНК вызвали тогда, в 70-е годы, широкий резонанс: передачи на радио, телевидении, статьи в научных и научно-популярных журналах.

### **6. Фазовые переходы в ян-теллеровских кристаллах**

Для изучения наиболее общих проявлений фазовых переходов необ-ходимо магнитные системы (в качестве модельных) дополнить ян-теллеровскими, удобными для изучения фазовых переходов 1-го рода и метастабильных состояний. Из работ Б.Я. Сухаревского и его сотрудников на примере гексагидрата фторсиликата меди следует, что такие системы имеют весьма богатую фазовую диаграмму: обнаружены три равновесные фазы последовательного упорядочения ян-теллеровских искажений, ряд метастабильных фаз, в том числе фаза, названная в этих работах «ян-теллеровским стеклокристаллом». Исследование этих фаз проведено комплексно методом низкотемпературной рентгенографии, ядерной гамма-резонансной спектроскопии, измерений теплоемкости и теоретического анализа — феноменологическая теория Ландау и теоретическая четырехподрешеточная модель. Наиболее важные фундаментальные результаты: проявление неустойчивости вблизи границы лабильности, связанной с возникновением ян-теллеровской динамики в упорядоченной фазе и переход между статическим и динамическим беспорядком.

## 7. Сверхпроводящие оксидные системы

Опыт исследования оксидной керамики, фазовых переходов в координационных кристаллах, изучение сверхпроводящих сплавов и ян-теллеровских систем успешно использован Сухаревским в работах по высокотемпературной сверхпроводимости в медьсодержащих сложных оксидах. В этих работах впервые показано, что сверхпроводимость реализуется в орторомбических фазах. Обнаружены изменения температурных зависимостей электросопротивления в нормальном состоянии и параметров решетки при структурных фазовых переходах и процессах упорядочения-разупорядочения в подсистеме «ионы кислорода–вакансии». Это позволило высказать гипотезу о роли ян-теллеровского упорядочения в механизме высокотемпературной сверхпроводимости.

В последние годы жизни усилия Б.Я. Сухаревского были сосредоточены на выяснении природы нормального и сверхпроводящего состояний таких своеобразных материалов, какими являются сложные медьсодержащие оксиды. В результате проведенной работы им сформулировано представление о сверхпроводящем кластере, структура которого близка к упорядоченной, а концентрация носителей обеспечивает заполнение целого числа зон. Эти условия определяют состояние носителей заряда на грани локализации, т.е. сильное электрон-ионное взаимодействие. Исследования Б.Я. Сухаревского и его учеников показали, что объем сверхпроводящей фазы равен объему сверхпроводящих кластеров. В последних его работах (1994–1995 г.г.) установлено, что температура сверхпроводящего перехода пропорциональна концентрации носителей в плоскостях $CuO_2$ соединений ВТСП. Эта концентрация соответствует границе устойчивости сверхпроводящей фазы.

## 8. Ячеистая наноструктура при образовании метаногидрата

При детальнейшем изучении диаграмм фазового состояния газогидрата метана (представителя нетрадиционных источников энергии) Б.Я. Сухаревским впервые было установлено существование новой промежуточной фазы, которая играет основную роль в образовании этого клатратного соединения: $CH_4 \cdot nH_2O$. Эта субстанция представляет собой мелкодисперсную двухфазовую смесь воды и метаногидрата, организованную в ячеистую мезоскопическую структуру. Размеры ячеек составляют ~$10^{-5}$ см и заполняют при изотермической выдержке весь объем экспериментальной ячейки. При T~274K толщина стенок метаногидрата составляет ~$10^{-7}$ см. Образование ячеистой наноструктуры обусловлено конкуренцией объемной (разность химических потенциалов, удельная энергия упругой деформации) и поверхностной энергий. При положительных температурах монолитный метаногидрат образуется при давлениях выше 6МПа.

Метаногидраты являются нетрадиционными источниками энергии, поскольку в процессе их разложения (при понижении давления) выделяются метан, а также вода.

Эти работы позволяют грамотно и эффективно проводить утилизацию выбросов метана из шахт, переводя его в метаногидрат, а также решать важную экологическую проблему по очистке атмосферы от выбросов $CH_4$.

В успешном решении указанных научных проблем наиболее активными сотрудниками Бориса Яковлевича Сухаревского были замечательные ученые-физики: Б. Алапин, И. Вишневский, А. Гавриш, С. Лысак (Харьковский институт огнеупоров), А. Алапина, Э. Андерс, А. Гуревич, И.Щёткин, В.Осика, Т.Пятигорская, В.Хоменко, И.Волчок (Физико-технический институт низких температур НАН Украины им. Б.И. Веркина), А. Быков, В. Ксенофонтов, В.Ганенко, Е. Цыбульский, Г. Шаталова, И. Жихарев, В. Васюков, А. Леонтьева, А. Прохоров, Г. Маринин, В. Коварский (Донецкий физико-технический институт НАН Украины им. А.А. Галкина).

Все перечисление научных достижений Б.Я. Сухаревского достаточно подробно характеризует его как одного из крупнейших ученых-физиков. Не менее важно отметить еще одну грань широты и всесторонности его интеллекта. Для Б.Я. не составляло труда разобраться в проблемах научных работ практически во всем спектре физики твердого тела. Его эрудицией пользовалось огромное число физиков, работавших в совершенно различных научных направлениях. Трудно оценить то огромное количество диссертаций, в которых Б.Я. выступал в качестве оппонента. Считалось, что пройти через «чистилище Сухаревского» — это гарантия и успешной защиты на Совете, и благополучного прохождения через ВАК. Хотя в процессе ознакомления Б.Я. с работой диссертантам приходилось иногда даже переделывать диссертацию, поскольку он находил зачастую новые важные аспекты, не замеченные диссертантом, или, наоборот, забраковывал целые разделы работы, а затем вместе с соискателем трудился над исправлениями. И всегда его выступления в качестве оппонента являлись украшением самого процесса защиты.

Завершая анализ научной деятельности профессора Б.Я. Сухаревского, следует отметить, что им подготовлен внушительный отряд замечательных ученых – двадцать кандидатов и два доктора физико-математических наук.
Труды Сухаревского и его учеников являются результатом целенаправлен-ных фундаментальных исследований, которые в ряде отмеченных выше случаев привели к конкретным практическим рекомендациям. Его работы (1953–1996 г.г.) способствовали формированию современных представлений о физике точечных дефектов в координационных кристаллах, их роли в теомодинамике и кинетике фазовых превращений, их проявлении в физических свойствах кристаллов. Эти результаты нашли прямое продолжение и в исследованиях высокотемпературной сверхпроводимости.

Б.Я Сухаревский был лидером в постановке научной проблемы, замечательным руководителем коллектива учёных, умел увлечь задачей и сплотить сотрудников для ее решения.

Будучи профессором университетов, отличался строгостью и красотой изложения материала, а также эмоциональностью, пробуждающей интерес студентов к теме лекции. Видимо, подражая К.Б. Толпыго, начинал лекцию прямо с порога, когда входил в аудиторию: «Так на чем мы остановились в прошлый раз?»

Многие годы Б.Я. Сухаревский был неизменным рецензентом журнала «Физика низких температур», причем, как правило, ему доставались самые трудные, зачастую, спорные работы.

Ученый совет ДонФТИ неоднократно выдвигал кандидатуру профессора Б.Я. Сухаревского для избрания в члены Академии наук Украины, однако, по причинам всем известным, ему всегда недоставало голосов для успеха...

Творческие достижения Бориса Яковлевича Сухаревского оставили яркий след в современной физической науке. И в памяти его коллег и учеников, друзей и сотрудников всегда будут живы воспоминания о присущем ему высоком профессионализме, замечательном чувстве юмора и неизменной доброжелательности в общении с людьми.

Его любимая шутка, обращенная к ученикам: «Такая получается историческая цепочка: вот вы — мои ученики, а я — ученик Б.Я. Пинеса, который считал себя учеником Я.И. Френкеля, в свою очередь учившегося у А.Ф. Иоффе. который, кажется, стажировался у Рентгена. Так что я - как бы звено, связывающее вас с нашими великими научными предками!»

Сегодня, в канун восьмидесятилетия профессора Бориса Яковлевича Сухаревского, хочется поблагодарить судьбу за встречу и многолетнее сотрудничество с таким замечательным Человеком и Ученым.

Академик НАНУ, В. Ерёменко, Харьков, ФТИНТ НАНУ им. Б.И. Веркина, Украина

Академик НАНУ, В. Манжелий, Харьков, ФТИНТ НАНУ им. Б.И. Веркина, Украина

Член-корреспондент НАНУ, В. Варюхин, Донецк, ДонФТИ им. А.А. Галкина, Украина

Член-корреспондент НАНУ, А.Алексеев, Донецк, ИФГП НАНУ, Украина

Профессор В. Белошенко, Донецк, ДонФТИ им. А.А. Галкина, Украина

Профессор А.Воронель, Тель Авив, Университет, Израиль

Профессор В. Пустовалов, Харьков, ФТИНТ НАНУ им. Б.И. Веркина, Украина

Профессор В.Малеев, Харьков, ИРЭ НАНУ им. А.Я. Усикова, Украина

Профессор С. Гредескул, Беер Шева, Университет им. Бен Гуриона, Израиль

Профессор Л. Межов-Деглин, ИФТТ РАН, Россия

Профессор А. Захаров, Новгород Великий, Университет, Россия

Профессор Н. Фогель, Хайфа, Технион, Израиль

Профессор И. Вишневский, Нью-Йорк, США

Профессор В. Цукерник, Реховот, Израиль

Профессор В. Васюков, Москва, ОАО "Газпром Промгаз", Россия
Профессор Э. Фельдман, Донецк, ИФГП НАНУ, Украина
Профессор А. Леонтьева, Хайфа, Израиль
Профессор Ю. Мамалуй, Донецк, Университет, Украина
Профессор Г. Левченко, Донецк, ДонФТИ им. А.А. Галкина, Украина
Профессор Ю. Медведев, Донецк, ДонФТИ им. А.А. Галкина, Украина
Профессор А.Д. Прохоров, Донецк, ДонФТИ им. А.А. Галкина, Украина
Профессор В. Юрченко, Донецк, ДонФТИ им. А.А. Галкина, Украина
Доктор Б. Алапин, Кобленц, Германия
Доктор В.Ксенофонтов, Майнц, Германия
Доктор А. Быков, Нью-Джерси, США
Доктор Е. Лакин, Хайфа, Технион, Израиль
Доктор Т. Пятигорская, Хайфа, Израиль
Доктор С. Лысак, Харьков, Украина
Доктор А. Эренбург, Беер Шева, Университет им. Бен Гуриона, Израиль
Доктор С. Коваленко, Харьков, ФТИНТ НАНУ им. Б.И. Веркина, Украина
Доктор Г. Шаталова, Донецк, ДонФТИ НАНУ им. А.А. Галкина, Украина
Доктор Е. Цыбульский, Донецк, ДонФТИ НАНУ им. А.А. Галкина, Украина
Доктор А. Ю. Прохоров, Донецк, ДонФТИ НАНУ им. А.А. Галкина, Украина
Доктор И. Жихарев, Луганск, ЛНУ имени Тараса Шевченко, Украина

| | СПИСОК ИЗБРАННЫХ СТАТЕЙ ПРОФ. Б.Я. СУХАРЕВСКОГО | | |
|---|---|---|---|
| | Название | Журнал | Соавторы |
| 1. | Исследование спекания прессованных порошков при всестороннем давлении | ЖЭТФ, 24, 1954, с.1613-1621 | Я.Е. Гегузин |
| 2. | Изменение и взаимодействие хромшпинелида с магнезиоферритом при нагревании | Докл. АН СССР, 109, 1956, с.1009 -1011 | Л.И. Карякин, П.Д. Пятикоп |
| 3. | О механизме образования и распада твёрдых растворов шпинелей и периклаза | Докл. АН СССР, 130, 1960, с.1095 -1098 | А.С. Френкель, К.М. Шмуклер и др. |
| 4. | О распаде твёрдых растворов в системе $ZrO_2$-CaO | Докл. АН СССР, 140, 1961, с.884 -887 | И.И. Вишневский |
| 5. | Об особенностях полиморфного превращения двуокиси циркония при охлаждении | Докл. АН СССР, 156, 1964, с.677-680 | Б.Г. Алапин, А.М. Гавриш |
| 6. | Влияние точечных дефектов на теплопроводность твёрдых растворов замещения | Физ. тверд. тела, 6, 1964, с.258-274 | И.И. Вишневский |
| 7. | Роль катионных вакансий при окислительно-восстановительных процессах в ионных кристаллах | Докл. АН СССР, 160, 1965, с.642 -645 | И.И. Вишневский |
| 8. | О бинарных твёрдых растворах замещения между неизоморфными кристаллами | Докл. АН СССР, 167, 1966, с.1046 -1049 | |

| № | Название | Источник | Авторы |
|---|---|---|---|
| 9. | О структурных вакансиях в двухкомпонентных ферритах-шпинелях | Докл. АН СССР, 171, 1966, с.359-362 | Б.Г. Алапин, Е.И. Аксельрод |
| 10. | Влияние скорости нагрева на температурные характеристики бездиффузионного превращения двуокиси циркония | Докл. АН СССР, 177, 1967, с.886-889 | А.М. Гавриш, П.П. Криворучко |
| 11. | Особенности температурной зависимости теплоёмкости ниобий-титанового сплава при переходе в сверхпроводящее состояние | ЖЭТФ, 54, 1968, с.1675-1679 | А.В. Алапина, Ю.А. Душечкин |
| 12. | Структурные вакансии и распад твёрдых растворов ферритов-шпинелей | J. Phys. Chem. Solids, 29, 1968, p.1773-1782 | Б.Г. Алапин, Е.И. Аксельрод |
| 13. | Термодинамика распределения катионных вакансий по подрешёткам в двухкомпонентных ферритах-шпинелях | Докл. АН СССР, 188, 1969, с.1045-1048 | Б.Г. Алапин, И.И. Вишневский |
| 14. | О влиянии неоднородности образцов на некоторые характеристики при фазовых переходах, близким к переходам второго рода | ЖЭТФ, 58, 1970, 1532-1542 | А.В. Алапина, Ю.А. Душечкин |
| 15. | Структурные и фазовые соотношения в нестехиометрических ферритах с недостатком кислорода | J. Phys. Chem. Solids, 32, 1971, p.1627-1639 | Б.Г. Алапин, Е.И. Аксельрод, И.И. Вишневский |
| 16. | Особенности температурных зависимостей коэффициентов теплопроводности твёрдых растворов системы ниобий-цирконий | Докл. АН СССР, 203, 1972, с.1044-1046 | Е.М. Савицкий, В.В. Барон, Э.Е. Андерс и др. |
| 17. | Определение границ флюоритоподобных твёрдых растворов в системе $HfO_2 - Y_2O_3$ | Докл. АН СССР, 208, 1973, с.1085-1088 | А.М. Гавриш, Е.И. Зоз, А.Е. Соловьёва |
| 18. | Диффузионные процессы в многоподрешёточных бинарных системах, содержащих структурные вакансии | Докл. АН СССР, 212, 1973, с.611-614 | И.И. Вишневский, Е.И. Аксельрод |
| 19. | Теплопроводность и температурная зависимость энергетических щелей образцов ниобия, содержащих значительное количество примесных атомов | ЖЭТФ, 64, 1973, с.1688-1698 | Э.Е. Андерс, И.В. Волчок |
| 20. | Термодинамика распада флюоритоподобных твёрдых растворов в системах на основе двуокиси циркония | Изв. АНСССР, Неорган. материалы, 11, 1975, с.465-470 | А.М. Гавриш, Е.И. Зоз |

| № | Название | Источник | Авторы |
|---|---|---|---|
| 21. | Кристаллизация фрагментов ДНК из водносолевых растворов | Докл. АН СССР, <u>224</u>, 1975, с.707-709 | В.Д. Осика, Т.Л. Пятигорская и др. |
| 22. | Энергетические щели и механизмы теплопереноса в сверхпроводниках с высоким содержанием примесных атомов | Физика Низких Температур, <u>2</u>, 1975, с.592-605 | Э.Е. Андерс |
| 23. | Исследование структуры и полиморфизма ортофосфата алюминия кристобалитового типа методами рентгенографии и инфракрасной спектроскопии | Кристаллография, <u>21</u>, 1976, с.322-326 | С.В. Лысак, Б.Г. Алапин |
| 24. | Структурные исследования молекулярных кристаллов ДНК | Studia Biophysica, <u>57</u>, 1978, p.35-36 | Б.И.Веркин, В.Д. Осика, Л.А.Фейгин |
| 25. | Исследование дегидратации корформационного состояния плёночных образцов ДНК методом дифференциальной сканирующей микрокалориметрии | Докл. АН СССР, <u>241</u>, 1978, с.703-706 | А.В. Алапина, С.Ю. Бидный |
| 26. | Влияние интеркаляции на низкотемпературную теплоёмкость слоистого кристалла иодистого свинца | Физика Низких Температур, <u>6</u>, 1980, с.933-938 | А.М. Гуревич, А.В. Алапина |
| 27. | Изменение анизотропии теплопроводности слоистого кристалла $PbY_2$ при одноосном сжатии | Физика Низких Температур, <u>7</u>, 1981, с.494-500 | Э.Е. Андерс, И.В. Волчок |
| 28. | Неустойчивость структуры $CuSiF_6 \cdot 6H_2O$ вблизи фазового перехода первого рода | Докл. АН СССР, <u>256</u>, 1981, с.1390-1393 | Ф. А. Бойко, А. М. Быков, В. Е. Ганенко и др. |
| 29. | Структура и термодинамические аспекты фазового перехода $Cu_{0.96}Zn_{0.04}SiF_6 \cdot 6H_2O$ | Кристаллография, <u>28</u>, 1983, с.488-494 | Е.О. Цыбульский, Г.Е. Шаталова |
| 30. | The Jahn-Teller Cooperative Effect in $Cu_{1-x}Zn_xSiF_6 \cdot 6H_2O$ | J. of Sold State Chemistry, <u>55</u>, 1984, p. 3092-3100 | F. A. Boyko, A. M. Bykov, V. E. Ganenko, *et al* |
| 31. | Предпереходные явления при структурных фазовых переходах первого рода | ЖЭТФ, <u>87</u>, 1984, с.1336-1348. | В.Г. Ксенофонтов, В.Л. Коварский и др. |
| 32. | Модель антиизоструктурного фазового перехода, связанного с упорядочением | Физика Низких Температур, <u>10</u>, 1984, с.1125-1131 | В.Л. Коварский, А.В. Леонтьева |
| 33. | Особенности внутреннего трения в твёрдом водороде | Физика Низких Температур, <u>11</u>, 1985, с.823-830 | Г.А. Маринин, А.В. Леонтьева и др. |
| 34. | Ян-Теллеровский стеклокристалл | Физика Низких Температур, <u>12</u>, 1986, с.1105-1108 | Ф. А. Бойко, Г.Ю Бочковая, А. М. Быков и др. |
| 35. | Структурные ян-теллеровские фазовые переходы, предшествующие сверхпроводящему переходу в керамических образцах $Ca_{2-x}Sr_xCuO_2$ | Письма в ЖЭТФ, 46, прилож., 1987, с.188-191 | Г.Е. Шаталова, В.Г. Ксенофонтов |
| 36. | Неустойчивость структуры Y-Ba-Cu-O вблизи перехода в сверхпроводящее состояние | Физика Низких Температур, 14, 1988, с.1108-1112 | Е.О. Цыбульский, Н. Е. Письменова, А.М. Быков и др. |
| 37. | Магнитное упорядочение и сверхпроводимость в La-Sr-Cu-O керамике с добавками железа | Физика Низких Температур, 14, 1988, с.540-543 | И.В. Вилкова, В.Г. Ксенофонтов, И.В. Рубан и др. |
| 38. | Проявление зарядового состояния сверхстехиометрического кислорода в мессбауровских спектрах лантан- стронциевой керамики | Физика Низких Температур, 14, 1988, с.1226-1230 | В.Н. Варюхин, И.В. Вилкова, Л.А. Ивченко и др. |
| 39. | Возможная роль конденсированного кислорода во внутреннем трении металлооксидных ВТСП | Физика Низких Температур, 15, 1989, с.992-994 | А.В. Леонтьева, Г.А. Маринин, В.М. Свистунов |
| 40. | Кристаллохимические особенности соединений ВТСП | Физика Низких Температур, 16, 1990, с.884-892 | |
| 41. | Изменение структуры и сверхпроводящих свойств при термоциклировании $YBa_3Cu_3O_4$ до 450°C | Кристаллография, 35, 1990, с.727-731 | Г.Е. Шаталова, С.И. Хохлова, И.В. Жихарев и др. |
| 42. | Проявление атомного упорядочения в характеристиках нормального и сверхпроводящего состояний ВТСП-оксидов $YBa_3Cu_3O_4$ | Физика Низких Температур, 17, 1991, с.971-986 | С.И. Хохлова, Г.Е. Шаталова, И.В. Жихарев и др. |
| 43. | Кислород в порах и внутреннее трение в высокотемпературных сверхпроводниках | Физика Низких Температур, 18, 1992, с.1-6 | А.В. Леонтьева, Г.А. Маринин, А.Ю. Прохоров, В.М. Свистунов и др. |
| 44. | Термодинамические особенности гидратообразования метана | Физ. Техн. Выс. Давл., 3, 1993, с.9-15 | В.Н. Васюков, А.В. Леонтьева, Ю.А. Макогон, А.Ю. Прохоров и др. |
| 45. | Аномалии низкочастотного внутреннего трения в кристаллическом метане | Физика Низких Температур, 20, 1994, с.815-820 | А.В. Леонтьева, Г.А. Маринин, А.Ю. Прохоров |
| 46. | Спектр вибронных состояний ян-теллеровского иона. | Физика Низких Температур, 20, 1994, с. 821-831 | В.Н. Васюков |
| 47. | Особенности термодинамики | Физика Низких | Э.А. Завадский |

| № | Название | Журнал | Авторы |
|---|---|---|---|
| | фазовых переходов в системе с двумя каналами упорядочения | Температур, 21, 1995, с.861-866 | |
| 48. | Взаимодействие d-электронов магнитного иона с локальными искажениями кристаллической структуры (обзор) | Физика Низких Температур, 21, 1995, с. 247-260 | В.Н. Васюков |
| 49. | Мезоскопическая структура гидрата метана. | Физ.Техн. Выс. Давл., 6, 1996, с. 64-68 | А.В. Леонтьева, П.М. Зоркий, В.Н. Васюков, А.Ю. Прохоров, Г.М. Алейникова, Б.С. Любарский, Б.А. Грядущий |
| 50. | Ячеистая наноструктура газогидратов | Журн. Структ. Хим., 38, 1997, с.867-874 | А.Ю. Прохоров, В.Н. Васюков, А.В. Леонтьева |
| 51. | Квазиаморфное состояние метаногидрата. | Журн.Структ.Химии, 39, 1998, с.86-91 | А.Ю. Прохоров, Б.Я. Сухаревский, В.Н. Васюков, А.В. Леонтьева |
| 52. | Ячеистая структура метаногидрата. Эксперимент и теория. | Химия в интересах устойчивого развития, 6, 1998, с.103-111 | А.Ю. Прохоров, В.Н. Васюков, А.В. Леонтьева, |
| 53. | Диаграмма фазовых состояний в YBa$_2$Cu$_3$O$_{6+\delta}$ при различных условиях атомного упорядочения | Физ.Техн. Выс. Давл., 12, 2002, с.54-58 | Б.Я. Сухаревский, И.В. Жихарев, С.И. Хохлова, Г.Е. Шаталова, С.Б. Бухтиярова |